\documentclass{ws-ijmpa}
\usepackage{rotating}
\begin{document}
\markboth{Dongsheng Du}
         {QCD Factorization and Rare $B$ Meson Decays}

%
\catchline{}{}{}{}{}
%

\title{QCD Factorization and Rare $B$ Meson Decays
       \footnote{Plenary talk given at the {\em International conference on 
        QCD and hadronic physics}, June, 16-20, 2005, Beijing, China}}
\author{\footnotesize DONGSHENG DU
        \footnote{Email:duds@mail.ihep.ac.cn}}
\address{Institute of High Energy Physics, Chinese Academy of Sciences,
       Beijing 100049, China \footnote{Mailing address}}
\maketitle


\begin{abstract}
Some recent progress in China in the study of charmless $B$ decays
with QCD factorization is reviewed. Chirally enhanced power
corrections and infrared divergence problem are stressed.
\end{abstract}
\section{Introduction}	
$B$ meson weak decays are very important for testing the standard
model and probing new physics. The two $B$ factories have accumulated
large data sets while Tevatron and LHCb will have even more data. The
main task for theorists is to compute all experimental observables in
a reliable way. This amounts to compute hadronic matrix elements
reliably. There are several methods in the literature. The earliest is
the so called navie factorization \cite{bsw}. In this method the
 matrix element has no renormalization scheme and scale dependence, so
 cannot cancel the corresponding dependence of the Wilson
 coefficients. Futher more, it cannot account for ``nonfactorizable''
 contributions. The generalized factorization \cite{ali} takes into
 account the one-loop radiative corrections to recover the
 renormalization scheme and scale dependence for the hadronic matrix
 elements. But at quark level, to avoid the infrared divergence, it
 has to assume the external quarks to be off-shell by $-p^{2}$. This
 results in gauge dependence. To account for ``nonfactorizable''
 contributions, a phenomenological parameter $N_{c}^{\rm eff}$ is
 introduced and it is assumed that $N_{c}^{\rm eff}$ is universal. But
 actually $N_{c}^{\rm eff}$ is process-dependent.

In order to overcome the shortcomings of the above methods, in 1999,
Beneke, {\em et al.} (BBNS) \cite{bbns} proposed a new scheme based on
QCD. In this talk, I will concentrate on our work \cite{ref4,ref5,ref6,ref7,ref8} 
about the chirally enhanced power
corrections and infrared divergence problem and the application of the
QCD factorization approach. My talk is organized as follows: In
Section 2, I shall give an simple introduction to QCD
factorization. Section 3 is devoted to the chirally enhanced power
corrections and the cancellation of the infrared divergence in vertex
corrections. In section 4${\sim}$6, the application of QCD
factorization including chirally enhanced power corrections to the
two-body charmless $B$-decays is reviewed. Section 7 is for conclusions.
\section{General remarks on QCD factorization}
Beneke, {\em et al.} (BBNS) proposed a new factorization scheme based
on QCD \cite{bbns}. They pointed out that, in the heavy quark limit
($m_{b}$ $\to$ $\infty$)
 {\small \begin{eqnarray}
  &&\!\!\!\!\!\!\!\!\!\!\!\!\!\!\!\!
  {\langle}M_{1}M_{2}{\vert}Q_{i}{\vert}B{\rangle}\ =\
   F_{i}^{B{\to}M_{2}}(q^{2}){\int}_{0}^{1}dx\,T_{i}^{I}(x){\Phi}_{M_{1}}(x)
  +(M_{1}{\leftrightarrow}M_{2}) \nonumber \\
  &&\!\!\!\!\!\!\!\! + \sum\limits_{j}{\int}_{0}^{1}d{\xi}\,dx\,dy\,
   T_{ij}^{II}({\xi},x,y){\Phi}_{B}(\xi){\Phi}_{M_{1}}(x){\Phi}_{M_{2}}(y)
  +{\cal O}({\Lambda}_{QCD}/m_{b})
 \label{eq:bbns}
 \end{eqnarray} }
where the short distance hard scattering kernels $T^{I,II}$ is
calculable order by order in perturbative theory; the long distance
quantities, e.g. decay constants, form factors, light-cone
distribution amplitudes are inputs form either experimental
measurements or other theory; $M_{1}$ is a light meson or a charmonium
state, $M_{2}$ (contains the spectator in $B$) is any light or heavy
meson. If $M_{2}$ is heavy (for example $D$), the second line in
Eq.(\ref{eq:bbns}) is $1/m_{b}$ suppressed.
\begin{figure}[h]
\begin{center}
\begin{picture}(300,100)(0,0)
\put(-50,-405){\epsfxsize150mm\epsfbox{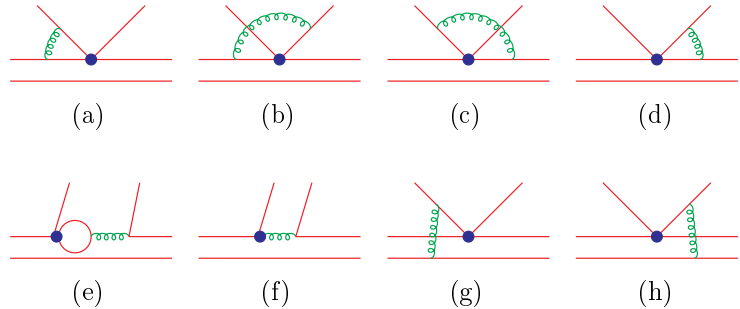}}
\end{picture}

\caption{\label{fig:fig1} $T^{I,II}$ at ${\cal O}({\alpha}_{s})$.
The upward quark lines denote the ejected meson $M_{1}$}
\end{center}
\end{figure}
At the zeroth order of ${\alpha}_{s}$, Eq.(\ref{eq:bbns}) reduces to
naive factorization. 
At the higer order, the corrections can be computed systematically.
The renormalization scheme and scale dependence of 
${\langle}Q_{i}{\rangle}$ is restored. In the heavy quark limit, the
``nonfactorization'' contributions is calculable perturbatively. It
does not need to introduce $N^{\rm eff}_{c}$.
The strong phase is suppressed by either ${\alpha}_{s}$ or $1/m_{b}$.
$W$-exchange and $W$-annihilation diagrams are $1/m_{b}$ suppressed. 
There is no long distance interactions between $M_{1}$ and ($BM_{2}$).
$T_{i}^{I}$ includes tree diagram, non-factorizable gluon exchange, e.g.
vertex corrections [Fig.\ref{fig:fig1}(a)-(d)], penguin corrections 
[Fig.\ref{fig:fig1}(e)-(f)], ${\cdots}$. $T_{ij}^{II}$ includes
hard spectator scattering [Fig.\ref{fig:fig1}(g)-(h)],
${\cdots}$ Later on we shall apply Eq.(\ref{eq:bbns}) to $B$ 
meson rare decays.
\section{Chirally Enhanced Power Corrections and Infrared Divergence
Cancellation in Vertex Corrections}
Superficially, ${\Lambda}_{QCD}/m_{b}$ ${\sim}$ $1/15$ is a small number.
But in some cases, such power suppression fails numerically. For example
  \begin{equation}
 {\langle}Q_{6}{\rangle}_{\rm F} = -2 \sum\limits_{q^{\prime}}
 {\langle}P_{1}{\vert}(\bar{q}q^{\prime})_{S-P}{\vert}0{\rangle}
 {\langle}P_{2}{\vert}({\bar{q}}^{\prime}b)_{S+P}{\vert}B{\rangle}
  \end{equation}
is always multiplied by a formally power suppressed but chirally enhanced
factor $r_{\chi}$ $=$ $2{\mu}_{P}/m_{b}$ (where ${\mu}_{P}$ $=$
$m_{P}^{2}/(m_{1}+m_{2})$, $m_{1,2}$ are current quark masses, and
$P_{1,2}$ denote pseudoscalar mesons).

In the heavy quark limit, $m_{b}$ ${\to}$ ${\infty}$, $r_{\chi}$ ${\to}$
$0$. But for $m_{b}$ ${\sim}$ 5GeV, $r_{\chi}$ ${\sim}$ ${\cal O}(1)$, no
$1/m_{b}$ suppression. So we should consider the chirally enhanced power
corrections. For example, in $B$ ${\to}$ ${\pi}K$,
dominant contribution to the amplitude ${\sim}$ $a_{4}+a_{6}r_{\chi}$, and
$a_{4}$ ${\sim}$ $a_{6}r_{\chi}$, so one half comes from the chirally
enhanced power corrections.

Possible sources of power corrections are : high twist wave functions,
quark transverse momentum $k_{\bot}$, and annihilation topology
diagrams. In this talk, I only discuss the chirally enhanced power
corrections from high twist wave functions \cite{ref4}. 

All the chirally enhanced power corrections involve twist-3 light cone
distribution amplitudes ${\Phi}_p(x)$ and ${\Phi}_{\sigma}(x)$. 
If we want to include chirally enhanced power corrections consistently,
we must prove that the hard scattering kernels are infrared finite, at
least for vertex corrections.

The decay amplitude of $B$ ${\to}$ $M_{1}M_{2}$ reads
 \begin{equation}
{\cal A}(B{\to}M_{1}M_{2})=\frac{G_{F}}{\sqrt{2}}\sum\limits_{q=\{u,c\}}
 \sum\limits_{i=1}v_{q}a^{q}_{i}({\mu}){\langle}M_{1}M_{2}{\vert}
 Q_{i}{\vert}B{\rangle}_{\rm F}
 \label{eq:amplitude}
 \end{equation}
For the hard kernel $T_{i}^{I}$ in Eq.(\ref{eq:bbns}), there is no
contributions from twist-3 wave function insertion for
$(V-A){\otimes}(V-A)$ and $(S+P){\otimes}(S-P)$ operators.
But for $(V+A){\otimes}(V-A)$, it is subtle!
For twist-3 light cone distribution amplitudes, if we use dimensional
regularization, the infrared divergences of Fig.1(a)-(d) can not cancel!
This is because the twist-3 wave functions are only defined in $4$ $=$
$3$ $+$ $1$ dimensions. We need to give gluon a small mass and
regularize the infrared integrals in 4-dimensions.

After a lengthy calculation, for the operator $Q_{5}$, the vertex 
correction terms are
 {\small \begin{eqnarray}&&\!\!\!\!\!\!\!\!\!\!\!\!
{\rm Fig.1(a)}\ {\sim}\ -\frac{{\alpha}_{s}}{4{\pi}}
 \frac{C_{F}}{N_{c}}\frac{{\Phi}_{\sigma}(v)}{v} \big[
 \frac{{\ln}^{2}{\lambda}}{2}+2{\ln}(-v){\ln}{\lambda}
 -4{\ln}v{\ln}{\lambda}+{\ln}{\lambda}+{\rm finite~terms} \big]
 \label{eq:q5a} \\ &&\!\!\!\!\!\!\!\!\!\!\!\!
{\rm Fig.1(b)}\ {\sim}\ +\frac{{\alpha}_{s}}{4{\pi}}
 \frac{C_{F}}{N_{c}}\frac{{\Phi}_{\sigma}(v)}{\bar{v}} \big[
 \frac{{\ln}^{2}{\lambda}}{2}+2{\ln}(-\bar{v}){\ln}{\lambda}
 -4{\ln}\bar{v}{\ln}{\lambda}+{\ln}{\lambda}+{\rm finite~terms}\big]
 \label{eq:q5b} \\ &&\!\!\!\!\!\!\!\!\!\!\!\!
{\rm Fig.1(c)}\ {\sim}\ +\frac{{\alpha}_{s}}{4{\pi}}
 \frac{C_{F}}{N_{c}}\frac{{\Phi}_{\sigma}(v)}{v} \big[
{\ln}^{2}{\lambda}-2{\ln}(-v){\ln}{\lambda}+3{\ln}{\lambda}
 +{\rm finite~terms} \big] \label{eq:q5c} \\ &&\!\!\!\!\!\!\!\!\!\!\!\!
{\rm Fig.1(d)}\ {\sim}\ -\frac{{\alpha}_{s}}{4{\pi}}
 \frac{C_{F}}{N_{c}}\frac{{\Phi}_{\sigma}(v)}{\bar{v}} \big[
{\ln}^{2}{\lambda}-2{\ln}(-\bar{v}){\ln}{\lambda}+3{\ln}{\lambda}
+{\rm finite~terms} \big] \label{eq:q5d}
 \end{eqnarray} }
where ${\lambda}$ $=$ $m_{g}^{2}/m_{b}^{2}$.

In the above, we can see that after summing over Fig.\ref{fig:fig1}(a)-(d),
the infrared divergences cancel only when ${\Phi}_{\sigma}(v)$ $=$
${\Phi}_{\sigma}(\bar{v})$, where $\bar{v}$ $=$ $1-v$. This is not a
surprise because we neglected the contribution of three-body twist-3 light
cone distribution amplitudes (LCDAs). For the asymptotic form,
${\Phi}_{\sigma}(v)$ $=$ $6v\bar{v}$ which is the same for both ${\pi}$
and $K$ meson.

After our work \cite{ref4}, BBNS \cite{ref9} also discussed the infrared
divergence problem. They implemented the equation of motion
\begin{equation}
 \bar{u}(k_{1})\not{k}_{1}{\cdots}{\upsilon}(k_{2})=0
 \label{eq:motion}
\end{equation}
to show the infrared safety without assuming ${\Phi}_{\sigma}(v)$ to
be symmetric. However Eq.(\ref{eq:motion}) is justified only when 2-particle
twist-3 LCDAs are asymptotic, i.e. ${\Phi}_{p}(v)$ $=$ $1$ and
${\Phi}_{\sigma}(v)$ $=$ $6v\bar{v}$ by neglecting the contribution from
the 3-particle twist-3 LCDAs. Our result is based on a more general
situation than that BBNS considered. Therefore, BBNS result is consistent
with ours.

We can prove at the order of ${\cal O}({\alpha}_{s})$ that the decay
amplitude is independent of the renormalization scale. We can also
prove the gauge invariance when chirally enhanced power corrections
are included. The detailed proof can be found in \cite{ref4}. We do
not discuss it here. Now we come to the application below.
\section{$B$ ${\to}$ $PP$, $PV$ Charmless Decays and $CP$ Violation}
Now we give the prediction of the branching ratios and $CP$ asymmetries.
We included the contributions of the hard spectator scattering and
annihilation topology. We also include the chirally enhanced power
corrections. We compute the branching ratios (${\cal B}r$) and $CP$ asymmetries of
$B$ ${\to}$ $PP$ : ${\pi}{\pi}$, ${\pi}K$, $K\bar{K}$,
       $K{\eta}^{(\prime)}$, ${\pi}{\eta}^{(\prime)}$,
       ${\eta}^{(\prime)}{\eta}^{(\prime)}$, ${\cdots}$ ${\cdots}$;
$B$ ${\to}$ $PV$ : ${\pi}{\rho}$, ${\pi}{\omega}$, ${\pi}K^{\ast}$
       ${\rho}K$, ${\omega}K$, $K^{\ast}{\eta}^{(\prime)}$,
       ${\rho}{\eta}^{(\prime)}$, ${\cdots}$ ${\cdots}$.
All our numerical results can be found in the tables in Refernce \cite{ref5,ref6}.
From our calculated resulte we see that :
Owing to including the chirally enhanced power corrections, the scale
${\mu}$ dependence of ${\cal B}r$ is smaller;
${\cal B}r$ of $B^{\pm}$ ${\to}$ ${\pi}^{0}$ ${\pi}^{-}$ (pure tree)
       and $B^{\pm}$ ${\to}$ $K^{0}{\pi}^{\pm}$ (pure penguin) are in good
       agreement with data;
${\cal B}r$ of $B^{0}$ ${\to}$ ${\pi}^{+}{\pi}^{-}$ is larger than data;
${\cal B}r$ of $B$ ${\to}$ ${\pi}K$ seem smaller than data;
For $B$ ${\to}$ $K{\eta}^{(\prime)}$, if consider di-gloun fussion,
       we can fit the data;
For $B^{0}$ ${\to}$ ${\pi}^{0}{\pi}^{0}$, our prediction is much smaller
       than data;
For $B^{0}$ ${\to}$ $K^{+}K^{-}$, the result is very small, only weak
       annihilation diagram contributes;
There are large uncertainties from CKM elements,form-factors,
annihilation parameters. But a global analysis can fit the data very well.

 For $CP$ asymmetries, the numberial results are not reliable because vertex,
 penguin, hard spectator scattering and annihilation diagrams can all give
 imaginary part to decay amplitudes, so strongly affect ${\cal A}_{CP}$.
 Because of space limitation, for more detail, see \cite{ref5}.
 For $B$ ${\to}$ $PV$, the differences from $B$ ${\to}$ $PP$ are:
 If the emitted meson is vector meson (vertex diagrams and penguins),
 then the twist-3 wave functions of $V$ do not contribute (power suppressed,
 so can be neglected).
 But for hard spectator scattering, even the emitted meson is vector one,
 there are still twist-3 contributions.
 Note, there are large uncertainties on $CP$ asymmetries. So we cannot make
 good predictions on ${\cal A}_{CP}$ in QCDF scheme.
 The calculated
 branching ratios of $B$ ${\to}$ $PV$ for $b$ ${\to}$ $d$ and $b$
 ${\to}$ $s$ transitions and $CP$ asymmetries
 can be found in Reference \cite{ref6}.
 \section{Global Analysis of $B$ ${\to}$ $PP$, $PV$ Charmless Decays}
 Beneke, {\em et al.} \cite{ref9} have done a global analysis on
 $B$ ${\to}$ $PP$. But their fitted ${\gamma}$ ${\sim}$ $90^{\circ}$,
 a bit large compare with the standard CKM fit.
 Now there are many new data on both $B$ ${\to}$ $PP$ and $PV$. It
 is necessary to do the global fit of $B$ ${\to}$ $PP$, $PV$ at the
 same time. We have done it!

 For $B$ ${\to}$ ${\pi}{\pi}$, ${\pi}K$, the fit is sensitive to
 ${\vert}V_{ub}{\vert}$, ${\gamma}\,({\rho},{\eta})$,
 $F^{B{\to}{\pi}}$, $F^{B{\to}K}$, $X_{A}$, $f_{B}/{\lambda}_{B}$,
 and $m_{s}$. To include seven $B$ ${\to}$ $PV$ decay channels, only
 $A_{0}^{B{\to}{\rho}}$, $X_{A}^{PV}$ are newly involved
 sensitive parameters. So the inclusion of $B$ ${\to}$ $PV$ will lead
 more stringent test to QCDF. 
 The experimental constrains of $CP$ asymmetries are not
 implemented, because the QCDF's predictions on $CP$ asymmetries
 in $B$ decays are rough!
 So we do not use $CP$ asymmetries for the global fit.
 The best fit values of the branching fractions is presented in Table.\ref{tab:tab8}.
 \begin{table}[h]
 \tbl{\label{tab:tab8} Fit1 and Fit2 mean the best fit values of ${\cal B}r$ in unit
 of $10^{-6}$ with
  and without the contribtuions of the chirally enhanced hard
  spectator and annihilation topology, respectively.}
 {\begin{tabular}{lccc|lccc} \hline
  \multicolumn{1}{c}{modes}  & Exp. & Fit1 & Fit2 &
  \multicolumn{1}{|c}{modes} & Exp. & Fit1 & Fit2 \\ \hline
         $B^{0}$ ${\to}$ ${\pi}^{+}{\pi}^{-}$
       & $4.77{\pm}0.54$ & 4.82 & 5.68 &
         $B^{+}$ ${\to}$ ${\pi}^{+}{\pi}^{0}$
       & $5.78{\pm}0.95$ & 5.35 & 3.25 \\
         $B^{0}$ ${\to}$ $K^{+}{\pi}^{-}$
       & $18.5{\pm}1.0$  & 19.0 & 18.8 &
         $B^{+}$ ${\to}$ $K^{+}{\pi}^{0}$
       & $12.7{\pm}1.2$  & 11.4 & 12.6 \\
         $B^{+}$ ${\to}$ $K^{0}{\pi}^{+}$
       & $18.1{\pm}1.7$  & 20.1 & 20.2 &
         $B^{0}$ ${\to}$ $K^{0}{\pi}^{0}$
       & $10.2{\pm}1.5$  & 8.2 & 7.3 \\
         $B^{+}$ ${\to}$ ${\eta}{\pi}^{+}$
       & $<5.2$          & 2.8 & 1.8 &
         $B^{0}$ ${\to}$ ${\pi}^{\pm}{\rho}^{\mp}$
       & $25.4{\pm}4.3$  & 26.7 & 29.5 \\
         $B^{+}$ ${\to}$ ${\pi}^{+}{\rho}^{0}$
       & $8.6{\pm}2.0$   & 8.9 & 8.5 &
         $B^{0}$ ${\to}$ $K^{+}{\rho}^{-}$
       & $13.1{\pm}4.7$  & 12.1 & 5.1 \\
         $B^{+}$ ${\to}$ ${\phi}K^{+}$
       & $8.9{\pm}1.0$   & 8.9 & 7.1 &
         $B^{0}$ ${\to}$ ${\phi}K^{0}$
       & $8.6{\pm}1.3$   & 8.4 & 6.7 \\
         $B^{+}$ ${\to}$ ${\eta}{\rho}^{+}$
       & $<6.2$          & 4.6 & 3.8 &
         $B^{0}$ ${\to}$ ${\omega}K^{0}$
       & $5.9{\pm}1.9$   & 6.3 & 1.2 \\ \hline
      \end{tabular} }
      \end{table} \\
 The main results are : \\
  The best fit of ${\gamma}$ ${\sim}$ $79^{\circ}$ which is
 consistent with recent fit results $37^{\circ}$ $<$ ${\gamma}$ $<$
 $80^{\circ}$\cite{ref12}. \\
 For $B$ ${\to}$ ${\pi}^{0}{\pi}^{0}$, the best fit is around
       $1{\times}10^{-6}$, \\ $~~~~~~~~$
       Exp: the {\sc BaBar} and Belle average is
       $(1.90{\pm}0.49){\times}10^{-6}$ \\
       For $B^{+}$ ${\to}$ ${\omega}K^{+}$, the best fit is
       $6.25{\times}10^{-6}$, \\ $~~~~~~~~$
       Exp: Belle $(9.2^{+2.6}_{-2.3}{\pm}1.0){\times}10^{-6}$ \ \ \
            CLEO  $<8{\times}10^{-6}$. \\
       For $B^{+}$ ${\to}$ ${\omega}{\pi}^{+}$, the best fit is
       $6.66{\times}10^{-6}$, \\ $~~~~~~~~$
       Exp: {\sc BaBar} $(6.6^{+2.1}_{-1.8}{\pm}0.7){\times}10^{-6}$ \ \ \
            Belle  $<8.2{\times}10^{-6}$ \\
       For $B^{+}$ ${\to}$ ${\pi}^{+}K^{{\ast}0}$, the best fit ${\sim}$
       $10{\times}10^{-6}$, \\ $~~~~~~~~$
       Exp: LP03 ${\sim}$ $(10.3{\pm}1.2^{+1.0}_{-2.7}){\times}10^{-6}$\ \ \
            ICHEP04 ${\sim}$ $(9.0{\pm}1.3){\times}10^{-6}$. \\
 For more details, see Reference \cite{ref7}.
\section{$B_{s}$ ${\to}$ $PP$, $PV$ Charmless Decays}
 We include : i) chirally enhanced power corrections,
              ii) weak annihilation,
              iii) hard spectator scattering.
We list the computed branching ratios and $CP$ asymmetries of
$B_{s}$ ${\to}$ $PP$, $PV$ in Reference \cite{ref8}.
Only $B_{s}$ ${\to}$ $K^{({\ast})}K$ $(10^{-6}{\sim}10^{-5})$,
 $K^{({\ast}){\pm}}{\pi}^{\mp}$  $(10^{-6})$,
 $K^{\pm}{\rho}^{\mp}$  $(10^{-5})$,
  ${\eta}^{(\prime)}{\eta}^{(\prime)}$ $(10^{-6}{\sim}10^{-5})$
 have large branching ratios.
 \section{Conclusions}
 We have shown that
 \begin{itemize}
 \item chirally enhanced power corrections must be included in QCDF
 \item twist-3 light cone distribution amplitude ${\Phi}_{p}(x)$,
       ${\Phi}_{\sigma}(x)$ must be considered simultaneously
 \item the infrared divergences in vertex corrections cancel only
       twist-3 wave function of light pseudoscalar is symmetric,
       so chirally enhanced power enhanced power corrections can
       be included consistently
 \item the calculated branching ratios for $B$ ${\to}$ $PP$, $PV$
       charmless decays are, principally, in agreement with data
 \item gloabl analysis of $B$ ${\to}$ $PP$, $PV$ can fit the data
       very well including $B^{0}$ ${\to}$ ${\pi}^{0}{\pi}^{0}$
       and ${\gamma}$ ${\sim}$ $79^{\circ}$
 \item $B_{s}$ ${\to}$ $PP$, $PV$ charmless decays are waiting to
       be tested in Tevatron and LHCb
 \end{itemize}
I thank H. J. Gong, J. F. Sun, D. S. Yang, and G. H. Zhu for their
collaborations and discussions.

\end{document}